\documentclass[%
preprint,
 amsmath,amssymb,
 aps, prl,
 showpacs,
longbibliography
]{revtex4-1}

\usepackage{graphicx}
\usepackage{epsfig}
\usepackage{rotate}
\usepackage{float}
\usepackage{amssymb}
\usepackage{amsbsy}
\usepackage{color}
\usepackage{overpic}
\usepackage{marvosym}
\usepackage{textcomp}
\usepackage{amsmath}
\usepackage{longtable}
\usepackage{graphicx}
\usepackage{dcolumn}
\usepackage{bm}

\newcommand\Nu{\text{Nu}}
\newcommand\Ra{\text{Ra}}
\newcommand\Rey{\text{Re}}
\newcommand\Pran{\text{Pr}}

\begin{document}
\preprint{APS/123-QED}

\title{Scaling relations in large-Prandtl-number natural thermal convection}

\author{Olga Shishkina}
 \email{Olga.Shishkina@ds.mpg.de}
\affiliation{Max Planck Institute for Dynamics and Self-Organization,
Am Fassberg 17, 37077 G\"ottingen, Germany}%
\author{Mohammad S. Emran}
\affiliation{Max Planck Institute for Dynamics and Self-Organization,
Am Fassberg 17, 37077 G\"ottingen, Germany }%
\author{Siegfried Grossmann}
\affiliation{Fachbereich Physik der Philipps-Universit\"at, Renthof 6, 35032 Marburg, Germany}%
\author{Detlef Lohse}
\affiliation{Physics of Fluids group, Department of Science and Engineering, Mesa+ Institute, 
Max Planck Center for Complex Fluid Dynamics
and J. M. Burgers Centre for Fluid Dynamics, University of Twente, P.O. Box 217, 7500 AE Enschede, The Netherlands}
\affiliation{Max Planck Institute for Dynamics and Self-Organization,
Am Fassberg 17, 37077 G\"ottingen, Germany}%

\date{\today}

\begin{abstract}
In this study we follow Grossmann and Lohse, {\it Phys. Rev. Lett.} {\bf 86}, 3316 (2001), 
who derived various scalings regimes for the dependence of the Nusselt number $\Nu$ and the Reynolds number $\Rey$ on the 
Rayleigh number $\Ra$ and the Prandtl number $\Pr$. 
We focus on theoretical arguments as well as on numerical simulations for 
the case of large-$\Pran$ natural thermal convection.
Based on an analysis of self-similarity of the boundary layer equations, we derive that in this case the 
limiting large-$\Pran$ boundary-layer dominated regime is I$_\infty^<$, introduced and defined in \cite{Grossmann2001},
with the scaling relations $\Nu\sim\Pran^0\,\Ra^{1/3}$ and $\Rey\sim\Pran^{-1}\,\Ra^{2/3}$. 
Our direct numerical simulations for $\Ra$ from $10^4$ to $10^9$ and $\Pr$ from 0.1 to 200 show that 
the regime I$_\infty^<$ is almost indistinguishable from the regime III$_\infty$, where the kinetic dissipation 
is bulk-dominated. With increasing $\Ra$, the scaling relations undergo a transition 
to those in IV$_u$ of reference \cite{Grossmann2001}, where the thermal dissipation is determined by its bulk contribution.
\end{abstract}



\maketitle

\section{Introduction}
Thermal convection flows are common in nature and technology.
One of the classical systems to study such flows is Rayleigh--B\'enard convection (RBC) 
\cite{Ahlers2009, Bodenschatz2000, Lohse2010, Chilla2012},
where a fluid is confined between a heated bottom plate and a cooled top plate.
In RBC the main input or control parameters of the system 
are the Rayleigh number $\Ra\equiv \alpha g \Delta H^3/(\kappa \nu)$,
the Prandtl number $\Pran\equiv\nu/\kappa$, and the cell geometry.
Here
$\nu$ denotes the kinematic viscosity, 
$\kappa$ the thermal diffusivity,
$\alpha$ the isobaric thermal expansion coefficient of the fluid,
$g$ the acceleration due to gravity,
$H$ the distance between the top and bottom plates, and
$\Delta\equiv T_+-T_->0$\, with 
$T_+$ and $T_-$ the temperatures of, respectively, the heated bottom and the cooled top plates.

The main global response characteristics of this convective system are 
the mean convective heat transport from bottom to top and the mean momentum transport by convection, which are 
represented, respectively, by the Nusselt number $\Nu$ and the Reynolds number $\Rey$.
How $\Nu$ and $\Rey$ depend on $\Ra$ and $\Pran$ is the main issue in investigations of thermally driven flows.

The kinetic and the thermal dissipation rates are fundamental concepts in turbulent thermal convection.
For some convective flow configurations, including RBC, it is possible to
derive analytical relations for the time- and volume-averaged kinetic and thermal dissipation rates $\epsilon_u$ 
and $\epsilon_\theta$ in terms of  $\Ra$ and $\Nu$ cf. \cite{Ahlers2009, Grossmann2000}.
For example, in RBC, it holds: 
\begin{eqnarray}
 \label{qqq1}
\epsilon_u&=&(\nu^3/H^4)(\Nu-1)\Ra\,\Pran^{-2},\\
 \label{qqq2}
\epsilon_\theta&=&(\kappa\Delta^2/H^2)\,\Nu.
\end{eqnarray}
Using these relations, Grossmann and Lohse developed a scaling theory (GL theory) 
\cite{Grossmann2000, Grossmann2001, Grossmann2002, Grossmann2003, Grossmann2004, Grossmann2011, Stevens2013}, 
which is based on a decomposition of $\epsilon_u$ and $\epsilon_\theta$
into their boundary-layer (BL) and bulk contributions.
Basically, the so-called scaling regimes I, II, III, and IV in the GL theory are associated with the BL--BL, bulk--BL, BL--bulk and bulk--bulk dominance in $\epsilon_u$ and $\epsilon_\theta$, respectively.
The assigned subscripts $u$ and $\ell$ to these regimes indicate the $u$pper-Pr and $\ell$ower-Pr cases, respectively.
Equating $\epsilon_u$ and $\epsilon_\theta$ to their estimated either bulk or BL contributions 
and employing the Prandtl--Blasius BL theory \cite{Prandtl1905, Blasius1908, Landau1987, Schlichting2000} for 
the thermal and viscous BL thicknesses, theoretically possible limiting scaling regimes followed.
In particular, for the case of large $\Pr$-number thermal convection, the various possible scaling regimes 
I$_u$, I$_\infty^<$, I$_\infty^>$, II$_u$, III$_u$, III$_\infty$, IV$_u$ were obtained, see Grossmann and Lohse 
\cite{Grossmann2001} for the details.

\begin{figure*}
\centering \includegraphics[width=0.45\textwidth]{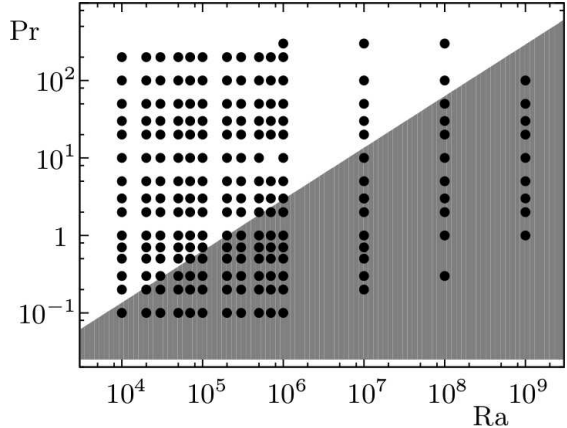}
\caption{Sketch of the conducted DNS (symbols) in an ($\Ra$, $\Pr$) plane.
White region corresponds to the scaling regimes I$_\infty^<$ and III$_\infty$.
Shadowed region corresponds to transition to other regimes.}
\label{figure1}
\end{figure*}

In this paper we first derive a scaling relation between $\Rey$, $\Nu$, and $\Pran$ in natural boundary-layer 
dominated thermal convection, in which the complete convective flow is due to temperature differences without any 
externally controlled wind input. This relation implies that in RBC the limiting 
large-$\Pran$ boundary-layer dominated regime is I$_\infty^<$,
with the scaling relations $\Nu\sim\Pran^0\,\Ra^{1/3}$ and $\Rey\sim\Pran^{-1}\,\Ra^{2/3}$. 
This regime I$_\infty^<$ matches the regime III$_\infty$ for larger $\Ra$, which in turn adjoins the regime IV$_u$ for 
even higher Rayleigh numbers. Based on the results of our direct numerical simulations (DNS) 
of RBC in a cylindrical container of aspect ratio 1, for $\Ra$ ranging from $10^4$ to $10^9$ and $\Pr$ from 0.1 to 200,
we demonstrate the correctness of the derived scaling relations in regime I$_\infty^<$ and also 
the transition to regime IV$_u$ for sufficiently large $\Ra$.

\section{Relation between $\Rey$, $\Nu$, and $\Pran$ in natural (purely thermally driven) boundary-layer dominated  thermal convection}

\begin{figure*}
\centering \includegraphics[width=\textwidth]{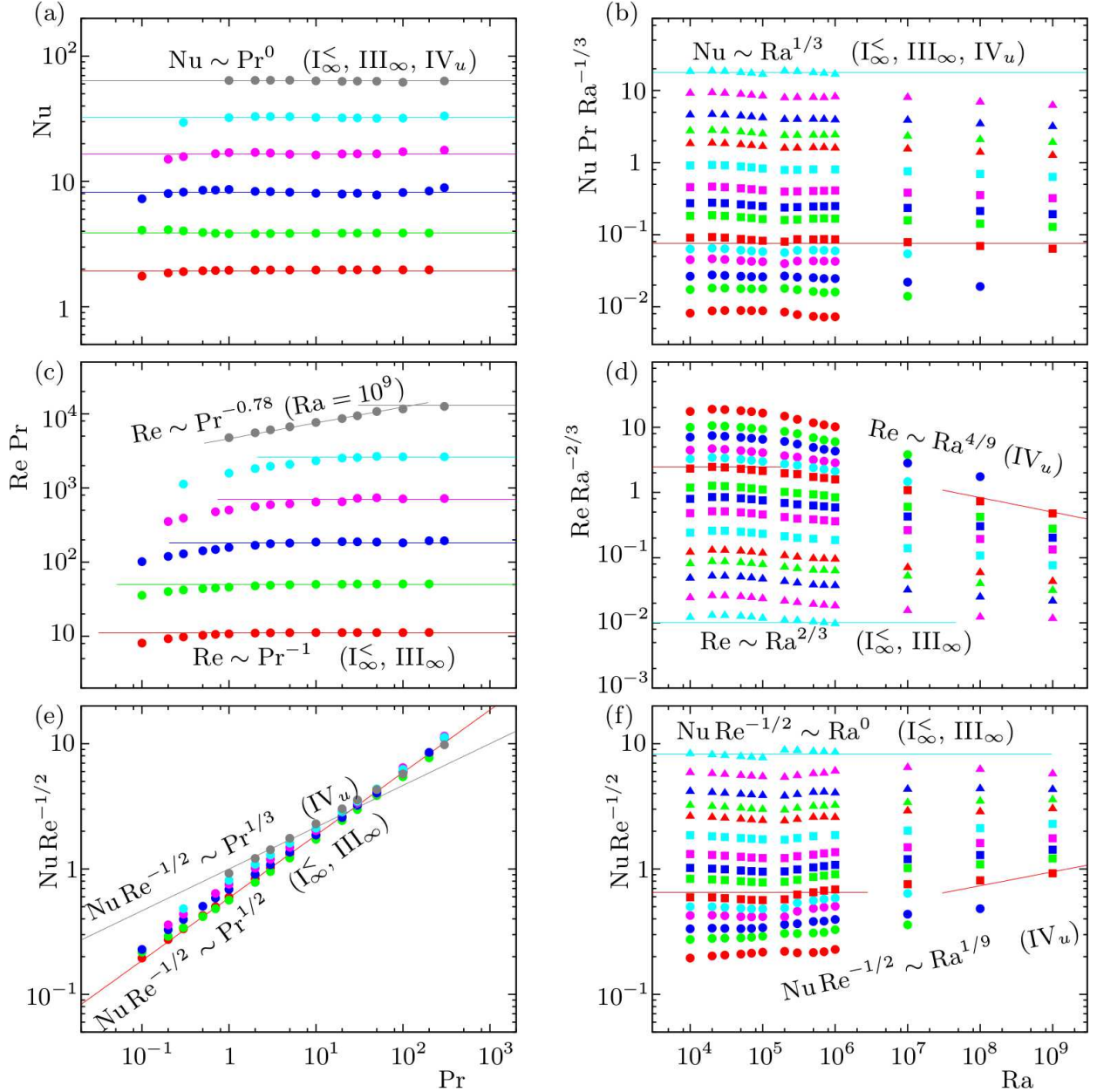}
\caption{
Nusselt number, Reynolds number and their combinations, as functions of the Prandtl number (a, c, e) and of the Rayleigh number (b, d, f), as obtained in the DNS of RBC in a cylindrical container of the aspect ratio 1,
for (a, c, e) $\Ra=10^4$ (red),  $\Ra=10^5$ (green),  $\Ra=10^6$ (blue),  $\Ra=10^7$ (magenta),  $\Ra=10^8$ (cyan) and  $\Ra=10^9$ (grey) and for (b, d, f) 
$\Pr=0.1$ (red circles),  $\Pr=0.2$ (green circles),  $\Pr=0.3$ (blue circles),  $\Pr=0.5$ (magenta circles),  $\Pr=0.7$ (cyan circles), 
$\Pr=1$ (red squares),  $\Pr=2$ (green squares),  $\Pr=3$ (blue squares),  $\Pr=5$ (magenta squares),  $\Pr=10$ (cyan squares), 
$\Pr=20$ (red triangles),  $\Pr=30$ (green triangles),  $\Pr=50$ (blue triangles),  $\Pr=100$ (magenta triangles),  $\Pr=200$ (cyan triangles). 
}
\label{figure2}
\end{figure*}

Following Prandtl \cite{Prandtl1905}, we consider a fluid flow along a 
heated plate and choose the coordinate system such that the $x$ direction is along the plate and 
the $z$ direction is pointing vertically away from the plate.
We assume that the mean flow in the other horizontal direction ($y$) is much weaker than that in $x$ or $z$ and, therefore,
consider a two-dimensional flow that depends on $x$ and $z$ only.
Under the standard Prandtl BL approximation \cite{Landau1987, Schlichting2000} we obtain the following 
momentum (\ref{mom1}) and thermal (\ref{en1}) BL equations
for a fluid motion near the isothermally heated horizontal plate: 
\begin{eqnarray}
\label{mom1}
 u_x\partial_xu_x+u_z\partial_zu_x&=&\nu\partial^2_zu_x+\beta\alpha g(T-T_0), \\
\label{en1}
 u_x\partial_xT+u_z\partial_zT&=&\kappa\partial^2_zT,
 \end{eqnarray}
where 
$T(x,z)$ denotes the temperature field above the plate,
$(u_x,u_z)$ is the velocity vector in the coordinates $(x,z)$  
and in the buoyancy term $\beta\alpha g(T-T_0)$ we have $\beta$ vanishing in the case of a horizontal heated plate 
and $\beta=1$ in the case of vertical heated plate.
The temperature boundary conditions are
\begin{eqnarray}
\label{bc1}
T(x,0)=T_+,&\quad& T(x,\infty)=T_0\equiv (T_++T_-)/2
 \end{eqnarray}
and the velocity vanishes at the plate,
\begin{eqnarray}
\label{bc2}
u_x(x,0)=u_z(x,0)=0,
 \end{eqnarray}
due to the assumed no-slip boundary conditions.
In the classical Prandtl--Blasius approach \cite{Prandtl1905, Blasius1908, Landau1987}, 
a constant flow above an infinite horizontal plate is considered, which is parallel to the plate.
This leads to the next boundary condition for the velocity, far away from the plate,
$u_x(x,\infty)=U\neq0$,
where $U$ is the so-called wind outside the BL.
The resulting Prandtl--Blasius approximation of the thermal boundary layer in laminar thermal convection 
is widely used, see, e.g. \cite{Grossmann2000, Zhou2010}.
This approach can be generalized to include the influence of the turbulent fluctuations within the BL \cite{Shishkina2015, Ching2017, Ching1997}
and also can be adapted to the case of a non-vanishing pressure gradient within the BL
(or to the case of a non-constant wind above the horizontal isothermal plate, which approaches the plate 
at arbitrary angle), see \cite{Shishkina2013, Shishkina2014}.
In all these cases the wind does not vanish in the bulk and, generally speaking, is artificially imposed 
in the boundary layer equations.

In natural thermal convection, where no mean flow is imposed in the core part of the domain, all the wind 
above the BL is self-organized. At the plate its absolute value is equal to zero.
With growing distance from the plate, the absolute value of the wind first increases and, after it 
has achieved its maximum, slowly vanishes back to zero.
The maximum value of the wind can be then considered as the reference quantity $U$.
The magnitude of $U$ and the location, where the maximum of the wind is achieved, depends implicitly on $\Ra$ and $\Pran$.
Such approach was considered for the case of vertical convection  
(between two isothermal differently heated vertical walls) in Shishkina \cite{Shishkina2016c}.
Here we use the ideas from \cite{Shishkina2016c}, to derive a relation between the Nusselt number $\Nu$ and the 
Reynolds number $\Rey$ in laminar natural convection for the case, where gravity is orthogonal to the 
isothermally heated plate, i.e., to the RBC case.

The idea of the method suggested in \cite{Shishkina2016c} is the following. 
Let us construct a similarity variable $\xi$ as the usual combination of the vertical and horizontal coordinates,
but now also with introduced Prandtl number and Rayleigh number:
\begin{eqnarray}
\label{sim1}
\xi&=&\Pran^a\Ra^b (x/L)^c (z/H).
\end{eqnarray}
Here $L$ is the length of the heated plate and the exponents $a$, $b$ and $c$ are not fixed for the time being.
In a similar way we introduce the exponents $d$, $e$, and $f$ for, respectively, $\Pran$, $\Ra$, and $x$ into 
the stream-function $\Psi$:
\begin{eqnarray}
\label{sim2}
\Psi&=&\nu\Pran^d\Ra^e(x/L)^f\phi(\xi),
\end{eqnarray}
from which the velocity components in horizontal and vertical directions can be evaluated as
$u_x=\partial_z\Psi$ and $u_z=-\partial_x\Psi$.
The temperature $T$ is represented though a non-dimensional temperature $\theta$ as follows:
\begin{eqnarray}
\label{sim3}
T&=&T_0+(T_+-T_0)\theta(\xi)=T_0+(\Delta/2)\theta(\xi).
\end{eqnarray}

The Nusselt number can then be calculated as
\begin{eqnarray}
\label{nu1}
\Nu\equiv\frac{-L^{-1}\int_0^L\left.\kappa\partial_zT\right|_{z=0}\,dx}{\kappa\Delta/H}=
\frac{-\theta'(0)}{2(c+1)}\Pran^a\Ra^b,
\end{eqnarray}
where $\theta'\equiv d\theta/d\xi$ is the derivative of $\theta$ with respect to the similarity variable $\xi$.
Analogously we can derive 
the Reynolds number, which is defined in terms of the maximal mean velocity 
along the heated plate:
\begin{eqnarray}
\label{re0}
\Rey\equiv UH/\nu, \quad
U\equiv\max_zL^{-1}\int_0^Lu_x\,dx.
\end{eqnarray}
Here the horizontal velocity $u_x$ equals 
\begin{eqnarray}
\label{ux1}
u_x=\partial_z\Psi=\Pran^{a+d}\Ra^{b+e}(\nu/H)(x/L)^{c+f}\phi'.
\end{eqnarray}
Assuming that the maximal velocity is achieved at a certain value of $\xi=\hat{\xi}$, where $\phi''(\hat{\xi})=0$,
from (\ref{re0}) and (\ref{ux1}) we obtain
\begin{eqnarray}
\label{re1}
\Rey=\Pran^{a+d}\Ra^{b+e}{(c+f+1)}^{-1}{\phi'(\hat{\xi})}.
\end{eqnarray}
This together with (\ref{nu1}) yields
\begin{eqnarray}
\label{rim1}
\frac{\Nu}{\Rey^{1/2}\,\Pran^{1/2}}=-\frac{\sqrt{c+f+1}}{2(c+1)}
\frac{\theta'(0)}{\;[\phi'(\hat{\xi})]^{1/2}}\Pran^{(a-d-1)/2}\Ra^{(b-e)/2}.
\end{eqnarray}
Now we substitute (\ref{sim1})--(\ref{sim3}) into the BL energy equation (\ref{en1}) and require the independence 
of the resulting BL equation from $\Pran$ and $\Ra$.
The equation then takes the following form:
\begin{eqnarray}
\label{en22}
\Pran^{a}\Ra^{b}(x/L)^{c}(1/H)\theta''\qquad\qquad\qquad\qquad\qquad\nonumber\\
+f\Pran^{d+1}\Ra^{e}(x/L)^{f-1}(1/L)\phi\,\theta'=0,
\end{eqnarray}
which implies
\begin{eqnarray}
\label{const1}
d=a-1,\quad e=b,\quad f=c+1.
\end{eqnarray}
The BL energy equation (\ref{en22}) then reduces to
\begin{eqnarray}
\label{en3}
\theta''
+\frac{c+1}{\Gamma}\phi\,\theta'=0
\end{eqnarray}
with the aspect ratio $\Gamma\equiv L/H$.
Note that the boundary conditions for $\phi$ and $\theta$ are also independent of $\Pran$ and $\Ra$,
i.e., $\phi(0)=\phi'(0)=0$, $\phi'(\infty)=0$, and $\theta(0)=1$, $\theta(\infty)=0$,
which one can obtain from (\ref{bc1})--(\ref{sim3}).

With (\ref{const1}), the relation (\ref{rim1}) is reduced to 
\begin{eqnarray}
\label{rim2}
\frac{\Nu}{\Rey^{1/2}\,\Pran^{1/2}}=-\frac{1}{\sqrt{2(c+1)}}
\frac{\theta'(0)}{\;[\phi'(\hat{\xi})]^{1/2}}\Pran^{0}\Ra^{0}.
\end{eqnarray}
The momentum equation (\ref{mom1}) with the similarity variable (\ref{sim1}) and the stream-function (\ref{sim2})
with the constants (\ref{const1}) takes the form
\begin{eqnarray}
\label{mom2}
\Pran^{4a}\phi_{\xi\xi\xi}&+&\Pran^{4a-1}[(c+1)\phi\phi_{\xi\xi}
-(2c+1)(\phi_\xi)^2]\nonumber\\
\label{mom2}
&-&\beta\,\Ra^{1-4b}(x/H)^{-4c-1}\theta/2=0.
\end{eqnarray}

With a proper choice of the constant $b$ ($b=1/4$ for $\beta\neq0$)
this equation can be released from the $\Ra$-dependence.
However, as one can see from the comparison of the first two terms in (\ref{mom2}), 
it cannot  be made generally independent of $\Pran$.
However, for $\Pran\gg1$, the second term in (\ref{mom2}) becomes negligible compared to the first term, which means that for $\Pran\rightarrow\infty$
there exist certain values of $a$, $b$, $c$, $d$, $e$, and $f$ in (\ref{sim1}) and (\ref{sim2}) such that 
the governing momentum (\ref{mom1}) and thermal (\ref{en1}) BL equations, written in terms of $\phi(\xi)$ and $\theta(\xi)$,  
as well as their boundary conditions, do not explicitly involve $\Pran$ and $\Ra$.
In this case, the solutions of the resulting BL equations, i.e., the functions $\phi(\xi)$
and $\theta(\xi)$, are also independent of $\Pran$ and $\Ra$.
In particular, the value of $\;{\theta'(0)}{\,[\phi'(\hat{\xi})]^{-1/2}}$ is a pure constant.
This together with (\ref{rim2}) immediately imply the following scaling relation
for the Nusselt number with the Reynolds and the Prandtl numbers, 
which thus in general holds for natural boundary-layer dominated large-Prandtl-number thermal convection:
\begin{eqnarray}
\label{scaling}
\Nu\sim\Rey^{1/2}\,\Pran^{1/2}.
\end{eqnarray}
Note that apart from RBC, the scaling (\ref{scaling}) was found also in other different configurations 
of natural large-Prandtl-number thermal convective flows, for example,
in horizontal convection, where the fluid layer is heated through one region of the bottom 
and cooled through another region of the bottom \cite{Shishkina2016, Shishkina2016a, Hughes2008}
and also in vertical convection, where the fluid is heated through one vertical surface of the fluid layer
and cooled though another vertical surface \cite{Shishkina2016c, Ng2015}.

Previously \cite{Grossmann2000} it was shown that the relation (\ref{scaling}) holds also for $\Pran\ll1$
in laminar Rayleigh--B\'enard convection.
Since the $\Pran$-dependence as in (\ref{scaling}) holds only for very large or small $\Pran$,
it formally breaks down for intermediate values of $\Pran$, which strictly speaking means the absence of the similarity solution.
However, also in this case relation (\ref{scaling}) provides a fair estimate due to the fact that
$\Pran\approx1$ in this regime and taking this value to some power does not introduce a strong $\Pran$-dependence.
Therefore, relation (\ref{scaling}) effectively and in good approximation holds for {\it all} Prandtl numbers, which 
is fully supported by our simulations as we will see in the next section.

\section{Scalings of $\Rey$ and $\Nu$ with $\Ra$ and $\Pran$ in boundary-layer dominated  large-$\Pran$ RBC (regime I$_\infty^<$)}

In order to obtain the second scaling relation, in addition to (\ref{scaling}), we follow 
\citet{Grossmann2000} for the BL-dominated thermal convection.
The balance of the time- and volume-averaged kinetic dissipation rate $\epsilon_u$
to its estimated BL contribution gives ${\epsilon_u}\sim(\nu U^2/\lambda_u^2)(\lambda_u/H)$,
where $\lambda_u$ is the thickness of the viscous BL near the bottom plate.
This together with Prandtl's relation $\lambda_u/H\sim\Rey^{-1/2}$ for laminar RBC flows \cite{Landau1987, Schlichting2000} yields
${\epsilon_u}\sim(\nu^3/H^4)\Rey^{5/2}$.
Thus, from (\ref{qqq1}), (\ref{scaling}) and 
the last relation one obtains the scalings in the laminar low-$\Pran$ regime I$_\ell$ of the GL theory:
\begin{eqnarray}
\label{Il1}
\Nu&\sim&\Pran^{1/8}\Ra^{1/4},\\
\label{Il2}
\Rey&\sim&\Pran^{-3/4}\Ra^{1/2}.
\end{eqnarray}
These scalings have been supported by numerous experimental and numerical RBC studies in the respective 
regions of the $(\Pran,\,\Ra)$ plane,
see, e.g., \cite{Kerr1996, Cioni1997, Verzicco1999, Verzicco2003, Poel2012, Stevens2013, Petschel2013};
therefore in the present work we do not focus on the low-$\Pran$ regime I$_\ell$.

With decreasing $\Ra$, the viscous BL thickness $\lambda_u$ generally increases and slowly saturates 
to a certain bounding value, which is comparable with $H$ \cite{Grossmann2001}.
In this case the BL contribution to the mean kinetic dissipation rate scales as
${\epsilon_u}\sim(\nu U^2/H^2)$, which yields
\begin{eqnarray}
\label{eu22}
{\epsilon_u}\sim(\nu^3/H^4)\Rey^{2}.
\end{eqnarray}
Thus, from (\ref{qqq1}), (\ref{scaling}) and  (\ref{eu22}) the scaling relations in the regime I$_\infty^<$
of BL-dominated large-$\Pran$ RBC follow,
\begin{eqnarray}
\label{Ilarge1}
\Nu&\sim&\Pran^{0}\,\Ra^{1/3},\\
\label{Ilarge2}
\Rey&\sim&\Pran^{-1}\,\Ra^{2/3}.
\end{eqnarray}
Note that these obtained scaling relations (\ref{Ilarge1}), (\ref{Ilarge2}) are similar to those 
in the regime III$_\infty$ \cite{Grossmann2001}, 
although in the regime III$_\infty$ the kinetic dissipation turns to be bulk-dominated.

The relations (\ref{Ilarge1}), (\ref{Ilarge2}) hold up to very large $\Pran$.
The DNS for $\Pran=2548$ and $\Ra$ up to $10^9$ \cite{Horn2013} showed that $\Nu\sim\Ra^{0.3}$ and $\Rey\sim\Ra^{0.6}$.
Independence of the Nusselt number of the Prandtl number for large $\Pran$ was demonstrated also in several other DNS, 
see, e.g. \cite{Poel2013, Petschel2013}.
The measurements by \citet{Xia2002} for $\Pran$ from 4 to 1350 and $\Ra$ up to $3\times 10^{10}$  
also demonstrated that $\Nu$ roughly goes as $\Ra^{0.3}$ and is almost independent of $\Pran$.

The results of our present DNS, which were conducted using the finite-volume code {\it goldfish} (see, e.g., \cite{Shishkina2015, Shishkina2016b}) 
for RBC for $\Ra$ from $10^4$ to $10^9$ and $\Pr$ from 0.1 to 200, are summarized in Fig.~\ref{figure2}.
In the left column (Fig.~\ref{figure2} a, c, e) the Prandtl number dependences 
and in the right column (Fig.~\ref{figure2} b, d, f) the Rayleigh number dependences are presented for the Nusselt number (Fig.~\ref{figure2} a, b),
Reynolds number (Fig.~\ref{figure2} c, d) and their combination $\Nu\,\Rey^{-1/2}$, due to the relation (\ref{scaling}).

One can see that through several decades of $\Pran$ and $\Ra$, the Nusselt number remains to be independent 
of $\Pran$ (Fig.~\ref{figure2}a) and scales with the Rayleigh number as $\sim\Ra^{1/3}$ (Fig.~\ref{figure2}b), 
in full agreement with (\ref{Ilarge1}).
Note that also for the larger $\Ra$ this scaling should hold, as on this end the regime   III$_\infty$
enters the regime IV$_u$ \cite{Grossmann2000, Grossmann2001} with its scaling relations:
\begin{eqnarray}
\label{I41}
\Nu&\sim&\Pran^{0}\,\Ra^{1/3},\\
\label{I42}
\Rey&\sim&\Pran^{-2/3}\,\Ra^{4/9}.
\end{eqnarray}
For the smallest $\Pran$ ($\Pran<0.5$ in Fig.~\ref{figure2}a), the Nusselt number  slightly grows with $\Pran$, 
as the flow undergoes a transition from the regime I$_\infty^<$ to the regime I$_\ell$, with its own scaling  (\ref{Il1}).

From Fig.~\ref{figure2}c we can conclude the following:
For sufficiently large $\Pran$ ($\Pran\geq 0.5$) and moderate $\Ra$ ($\Ra<10^9$), the Reynolds number scales as $\Rey\sim\Pran^{-1}$ through several decades of $\Pran$,
as it should be in the regimes I$_\infty^<$ and III$_\infty$, see (\ref{Ilarge2}).
Again, the region of very small $\Pran$, $\Pran<0.5$, belongs to the scaling regime I$_\ell$ (\ref{Il2}) and, therefore, 
the values of ($\Rey\,\Pran$) increase with increasing $\Pran$.
The large-$\Ra$ region is already around the transition to the regime IV$_u$, where the scaling (\ref{I42}) should take over.
For $\Ra=10^9$, the Reynolds number behaves already as $\Rey\sim\Pran^{-0.78}$, as our simulations show. 
Fig.~\ref{figure2}d also supports the scaling (\ref{Ilarge2}), but this time with respect to the $\Ra$-scaling in 
large-$\Pran$ BL-dominated RBC. Indeed, $\Rey$ goes as $\sim \Ra^{2/3}$ there, with a tendency to $\sim \Ra^{4/9}$ (\ref{I42}),
as it should be in the scaling regime IV$_u$ (this slope is shown in  Fig.~\ref{figure2}d with a red inclined line).

Finally, in Fig.~\ref{figure2}f one can see that $\Nu\,\Rey^{-1/2}$ in the regimes I$_\infty^<$ and III$_\infty$
is independent of $\Ra$. For larger $\Ra$, this quantity starts to increase and tends to $\Nu\,\Rey^{-1/2}\sim\Ra^{1/9}$,
as it should be in the scaling regime IV$_u$ (this slope is shown in  Fig.~\ref{figure2}f with a red inclined line).
The Prandtl-number dependences of $\Nu\,\Rey^{-1/2}$ in Fig.~\ref{figure2}e also support the scaling relations 
(\ref{Ilarge1}), (\ref{Ilarge2}) for the regimes I$_\infty^<$ and III$_\infty$
and the scaling relations (\ref{I41}), (\ref{I42}) for the regime IV$_u$, i.e.
it varies from $\Nu\,\Rey^{-1/2}\sim\Pran^{1/2}$ for smaller $\Ra$ to $\Nu\,\Rey^{-1/2}\sim\Pran^{1/3}$ for larger $\Ra$.

\section{Conclusions}

In the present work we derived that the relation $\Nu\sim\Rey^{1/2}\Pr^{1/2}$ 
holds in laminar natural thermal convection, where no wind is imposed above the isothermal plate,
for large Prandtl numbers $\Pran\gg1$.
Our derivation is based on a generalization of the Prandtl approach \cite{Prandtl1905} and 
on a search of a similarity solution  of the laminar thermal BL equation when $\Pran\rightarrow\infty$.
The scaling relation (\ref{scaling}), $\Nu\sim\Rey^{1/2}\Pr^{1/2}$,  which holds for all Prandtl numbers 
in laminar Rayleigh--B\'enard convection, holds more generally also in the non-laminar regimes.
It strictly holds for very large or very small $\Pran$, but formally breaks down for intermediate $\Pran$.
However, as in this case $\Pran\sim1$, the relation $\Nu\sim\Rey^{1/2}\Pr^{1/2}$ still provides a good approximation of the relationship between
$\Rey$, $\Pran$ and $\Nu$ also in this regime, which is fully supported by our simulations in a wide parameter range.

Because of the relation (\ref{scaling}), the limiting large-$\Pr$ laminar regime in RBC is regime I$_\infty^<$ 
with the scaling relations (\ref{Ilarge1}), (\ref{Ilarge2}), 
which were originally derived in \cite{Grossmann2001} as one of the 
possible scaling regimes in large-$\Pran$ thermal convection.

Based on our DNS data for $\Ra$ from $10^4$ to $10^9$ and $\Pran$ from 0.1 to 200 (totally $200$ cases), we 
showed that the scaling regime I$_\infty^<$ undergoes a smooth transition into regime III$_\infty$, 
so that one does not necessarily have to distinguish them.
For sufficiently large $\Ra$, the scaling relations in regime III$_\infty$ undergo a transition to those of regime IV$_u$.
All these scaling relations and transitions have been supported by our DNS over a large range of $\Ra$ and $\Pran$.

\begin{acknowledgments}
OS acknowledges financial support the German Research Foundation (DFG) under the grants Sh405/3-2 
and Sh405/4-2 (Heisenberg fellowship) and the Leibniz Supercomputing Centre (LRZ) for providing computing time.
We also acknowledge support by the Netherlands Organisation for Scientific Research (NWO) and the Max Planck Center for Complex Fluid Dynamics.
\end{acknowledgments}


%

\end{document}